\documentstyle[emulateapj,onecolfloat,psfig]{article}   

\newlength{\tskip}
\newlength{\colwidth}

\newcommand{\beq}{\begin{equation}}
\newcommand{\eeq}{\end{equation}}
\newcommand{\beqa}{\begin{eqnarray}}
\newcommand{\eeqa}{\end{eqnarray}}

\begin{document}
\twocolumn[   

\title{Kuiper Belt and Oort Cloud Objects: Microlenses or Stellar Occulters?}
\author{Asantha Cooray\altaffilmark{1}}
\affil{
Division of Physics, Mathematics and Astronomy, California
Institute of Technology, Pasadena, CA 91125.\\
E-mail: asante@caltech.edu}

\begin{abstract}
The occultation of background stars by foreground Solar system objects, such as planets and asteroids,
has been widely used as an observational probe to study physical properties associated with 
the foreground sample.  Similarly, the gravitational microlensing of background
stellar sources by foreground mass concentrations has also been widely used to understand the 
foreground mass distribution. Though distinct, these two possibilities present two extreme cases 
during a transit; At the edge of the Solar system and beyond, 
the Kuiper belt and Oort cloud populations may present 
interesting foreground samples where combinations of occultation and lensing, and possibly both during
the same transit, can be observed. To detect these events, wide-field 
monitoring campaigns with time sampling intervals of order tenths of seconds are required.
For certain planetary occultation light curves, such as those 
involving Pluto, an accounting of the gravitational lensing effect may be necessary when 
deriving precise physical properties of the atmosphere through the associated refraction signal.
\end{abstract}

]

\altaffiltext{1}{Sherman Fairchild Senior Research Fellow} 

\section{Introduction}

The occultation of background stars by foreground objects has been widely used as an observational probe to
study physical properties of various Solar system constituents such as planets, 
asteroids, comets, and rings (for an early review, see, \cite{Ell79} 1979).  
In addition to basic physical properties, such as the 
radius of the foreground source that occulted the background star, the refraction of background stellar 
light by the foreground planetary atmosphere
provides a well utilized probe to derive certain physical properties of the lower atmosphere
(see, for example, the recent review by \cite{EllOlk96}  1996 for further details). 

On the other hand, at galactic distance scales well beyond the Solar system, 
foreground mass concentrations are expected to gravitationally microlens background stars 
(\cite{Pac86} 1986). The microlensing of background sources by foreground objects is now well
utilized to understand mass distributions in the galaxy, including
potential dark matter candidates involving the so-called 
Massive Compact Halo Objects (MACHOs; \cite{Gri91} 1991). A typical microlensing observational campaign now involve continuous monitoring 
of million or more stellar sources
 towards, say, the galactic bulge and the Magellanic clouds with 
time sampling intervals of order tens of 
minutes or more (e.g., \cite{Alcetal93} 1993; \cite{Udaetal93} 1993).

Though occultation, with a decrease in background source flux, and microlensing, with an
increase in background source flux, have been mostly discussed as two separate phenomena, 
the two possibilities form essentially extreme cases during the transit of a foreground source
across the surface of a projected background stellar surface. 
In general, though, one expects signatures of both occultation and gravitational lensing
to be evident for a given population of foreground sources, 
the nature provides a simple reason why only these two extreme 
cases have been observed so far; the distance scale 
involved is such that known objects in the Solar
system always occult background sources while foreground sources at galactic distances 
always microlens background stars. 
The transition between that of an occultation to a lensing event is rapid
with only a limited range of parameters where both an occultation 
and a lensing signature will be
visible during the same transit.

The favorable condition to observe both a combination of occultation and microlens in the same 
foreground sample involve a 
projected extent to the foreground object that is of the same order as the Einstein radius 
associated with gravitational lensing. 
In the case of a pure occultation, the projected foreground source radius is 
larger than the Einstein radius while the opposite is true for the observation of a gravitational  lensing effect. While transit events involving binary stars have been previously suggested
as potential occurrences of both microlensing and occultation (\cite{Mar01} 2001), 
favorable conditions may also be present with foreground sources in the Solar system, but
at distance scales well beyond planets. 

Here, we identify  outer Solar system populations, such as 
the Kuiper belt objects (KBO; \cite{Kui51} 1951) and
the Oort cloud objects (OCO; \cite{Oor50} 1950)
as interesting samples of foreground sources where both
gravitational lensing and occultation, as well as a combination of the two, 
can be observed when they transit background stars. 
In fact, KBOs have been suggested as potential occulters in a previous study  where the use
of transits was considered in detail to extract the small size members of this population
(Roques \& Moncuquet 2000). Here, we suggest that the consideration of KBOs as occulters 
may only apply to the currently observed
KBO object population. If KBOs extend to much larger distances, as far as the Oort cloud, and
contains massive members at large distances that currently probed,
then there is some possibility that the distant
members may in fact produce either a signature of lensing alone or a combination of lensing and
 occultation. 

While no detailed data on the OCO population are observationally available,
the currently cataloged KBO population is mostly at orbital distances between 
40 and 50 AU with over $10^5$ objects of 100 km or more in size.
The total estimated mass is of 
order 0.08 $M_{\earth}$ (see, \cite{LuuJew02} 2002 for a recent review).
Extending current wide-field microlensing campaigns, which have been well executed to monitor
millions of stars or more on a given night  (e.g., \cite{Alcetal93} 1993; \cite{Udaetal93} 1993), we suggest that small bodies
of the outer Solar system can be detected and cataloged via similar continuous monitoring 
programs. Since the duration of transit events involving outer Solar system objects 
are of order a minute and less,
sampling time intervals, however, must be at the level of few tenths of seconds instead of
usual tens of minutes or more  time scales currently used in galactic microlensing campaigns.
Such high sampling rates, while keeping the same flux threshold levels as current surveys, are
within reach with the advent of dedicated large area telescopes 
and continuous improvements in the instrumental front.
The combined occultation and lensing measurements allow the observational data on
KBO and OCO samples to be significantly extended since one is no longer sensitive to 
individual fluxes, as in direct imaging observations,
but rather on the ability to detect and extract transit events
during the monitoring of a large sample of background sources.

The discussion is organized as following. In the next section, we introduce the concept of 
occultation  and microlensing
as two extreme cases of the same transit event. We discuss potential signature of KBOs 
when transiting across a
background stellar surface. A detailed study of the statistical signature of 
occultation due to KBO population 
is presented in Roques \& Moncuquet (2000) to which we refer the reader to further details. Similarly,
the paper by Agol (2002) considers the
signature of an occultation and lensing during transit events 
that involve binary sources orbiting
each other. A prior discussion on the occultation 
signature in a microlensing light curve, as applied to
 galactic lensing surveys, is available in \cite{Bro96} (1996).
Here, we consider the application to the Solar system as a potential way to 
extend our understanding of outer members which may 
have avoided direct detection due to low flux levels.

\section{Occultation and Gravitational lensing}

Following \cite{BarNar99} (1999), we write 
the ``lens equation'' associated with a gravitational lensing event as
\begin{equation}
\beta = \theta - \frac{\theta_E^2}{\theta} \,,
\end{equation}
where $\theta$ corresponds to image positions given the source position, $\beta$. Here, $\theta_E$ is the
angular Einstein radius for a foreground source of mass $M$ given by
\begin{eqnarray}
\theta_E&=& \sqrt{\frac{4 GM}{c^2}\frac{D_{ls}}{D_s D_l}} \, ,\nonumber \\
	&\approx& (7.1 {\rm mas}) \left(\frac{M}{M_{\earth}}\right)^{1/2} \left(\frac{D_l}{100 \rm AU}\right)^{-1/2} \, ,
\end{eqnarray}
where we have made use of the fact that the distance between foreground lens and background source, $D_{ls}=D_s-D_l$, can be well approximated by source distance, $D_s$, with the final result only depending on the
foreground lens distance, $D_l$. For a Earth mass object at a distance of 100 AU, the
 Einstein radius, $R_E=\Theta_E D_l$, is of order $5.1 \times 10^2$ km $(M/M_{\earth})^{1/2} (D_l/100 \rm AU)^{1/2}$. Foreground objects whose radius, $R_L$, is equal or less than the Einstein radius are expected to microlens background sources during a transit event.

\begin{figure}[t]
\centerline{\psfig{file=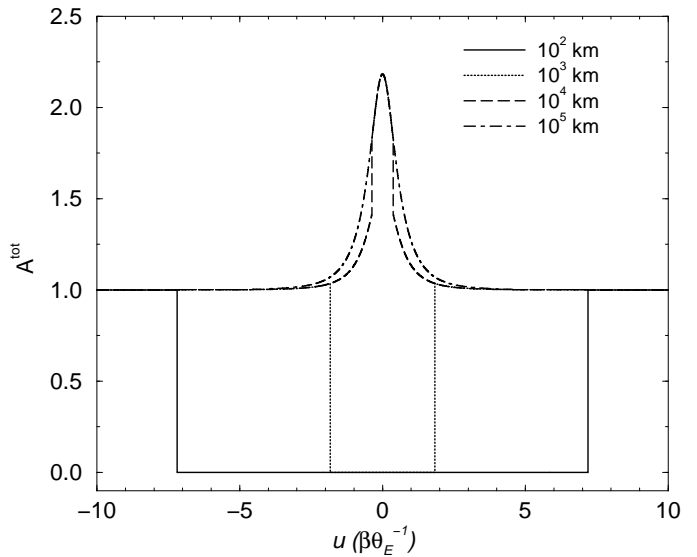,width=3.5in,angle=-90}}
\caption{The total magnification, $A^{\rm tot}$, during a transit of an outer Solar system object across
a background stellar object as a function $u=\beta \theta_E^{-1}$ 
(the angular separation of background point stellar source
from the foreground object in units of the Einstein angle). 
We assume a foreground lens distance, $D_l$, of $10^4$ AU consistent
with distance to the Oort cloud. We show several possibilities as a function of the object radius
ranging from 10$^2$ km to 10$^5$ km. At higher radii, a lensing event is clearly visible while at lower
radii, an occultation is present. We assume a density for the objects consistent with that of Pluto.}
\label{fig:lc}
\end{figure}

Under the point mass approximation, the microlensing event involve two images.
The total magnification during the transit is given by the sum of 
individual magnifications with a
correction that accounts for the potential occultation by the foreground source 
(for example, Agol 2002):
\begin{equation}
A^{\rm tot} = A_{-} \Theta(f_1-f_2) +A_{+} \Theta(f_1+f_2) \, ,
\label{eqn:atot}
\end{equation}
where $A_{-}$ and $A_{+}$ are magnifications associated with inner and outer 
images, which  are identified  with respect to the Einstein radius. Defining the projected
lens-source angular separation in terms of the Einstein angle, $u=\beta \theta_E^{-1}$, we write
\begin{eqnarray}
A_{\pm} = \frac{1}{2}\left(\frac{u^2+2}{u\sqrt{u^2+4}} \pm 1\right) \, .
\end{eqnarray}
In equation~\ref{eqn:atot}, $\Theta(x)$ is the step function of $x$ with $f_1$ and $f_2$ given by
\begin{eqnarray}
f_1 &=& 1.0-\left(\frac{R_L}{R_E}\right)^2 \; {\rm and} \nonumber \\
f_2 &=& \left(\frac{R_L}{R_E}\right)u \, ,
\end{eqnarray}
respectively.
As written, if $R_L > R_E$, the inner image ($-$) is occulted while the outer image (+) 
is also occulted when $u < R_L/R_E-R_E/R_l$.  The general condition for an occultation requires
that both images are occulted at all times. This depends on the impact distance of the transit
chord with respect to foreground lens center. In the case of a transit event that passes exactly
through the center, the condition is $\sqrt{2} R_L > R_E$.

We illustrate several occultation light curves,
with a lens source distance corresponding to the OCO population in Fig.~1. As shown, the light curves
exhibit the transition from a full occultation to that of a lensing event with increasing radius, or mass, for the foreground object. For simplicity, here and throughout, we assume a density for 
the object equivalent to that
of Pluto. In Fig.~1, we plot these light curves in terms of the
angular separation, in units of Einstein angle, between the background stellar source and the 
foreground object. One can convert these light curves to a more practical unit, such as time,
based on information related to the projected relative 
velocity of the foreground object across the background source as viewed from Earth. Assuming
typical velocities of order 30 km sec$^{-1}$ near the ecliptic, we determine event durations of
order $\sim$ 3 sec to  3000 sec for radii ranging from 100 km to 10000 km. In reality, members of the
outer Solar system objects  are likely to have radii that are less than 1000 km, or possibly even
less than 100 km, suggesting that time durations are likely to be at the order of 10 seconds or less.

In estimating this event time duration,
we have assumed that the background source radius is much less than that of the foreground
object.  In the event that the projected background source radius at the distance of the
foreground object is larger than that of the lens, the time duration will be increased from that
of the foreground source size to background source size. The actual flux
ratio measured during the transit will also be determined by a combination of 
the flux of the background source, the flux of the foreground lens, and their projected radii at the foreground source distance.
Since one expects outer Solar system to be significantly faint, 
if the background stellar radius is smaller than the source, then the fractional 
flux variation can be significant; 
In the case of an occultation, the effect can be a complete decrease in the background stellar flux.  If the projected background stellar size
is more than that of the foreground object, then the fractional flux difference during an
occultation would be equivalent to the ratio of planet to star projected areas.
This variation, however, can be tens of percent or more and not likely to be always at 
the few percent level or less expected for transit
events that involve extra-Solar planets that occult their central stars.  In the case of
a lensing event, the flux increase depends on the detailed properties of the transit such as the
minimum impact distance of the background source relative to foreground object center. For realistic
scenarios involving minimum impact distances of order the Einstein radius, we expect 
fractional flux variations, in terms of an increase now, to be also of order 100\%.

Thus, the limiting factor related to the
detection of outer Solar system transit events is not the precision of
relative photometry, as in the case of extra-Solar planetary transit searches, but rather the 
time separation of successive images. While current wide-field microlensing and extra-Solar
 transit
surveys do not obtain data with time samplings at few tenths of seconds or less, 
for the purpose of understanding the outer Solar system, one will need
imaging data separated at small time intervals. 
Note that high data sampling rates, corresponding to
few hundreds of micro seconds, are already possible towards small sky areas with 
unique occultation CCD imagers developed by various groups that observe planetary occultations in 
the Solar system.
With dedicated large area telescopes, and further developments in the experimental front,
it is likely that in the near future millions of stars can be monitored on the night sky 
with data taken at sub second intervals.

\begin{figure}[t]
\centerline{\psfig{file=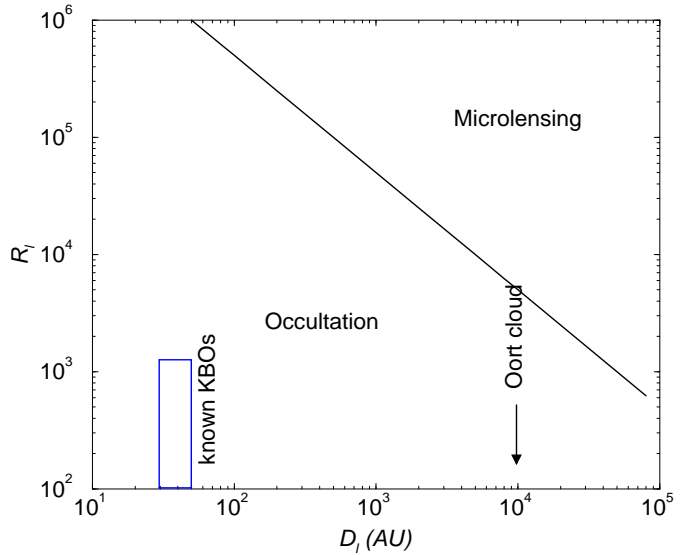,width=3.5in,angle=-90}}
\caption{The occultation vs,. microlensing possibilities in the outer solar system. Here, we plot
the radius of the foreground object as a function of the distance, $D_l$.
The solid curve shows an estimated division between occultation and microlensing possibilities
which we take to be the case when $R_L \sim R_E$. For comparison, we show the known KBO population
as well as the suggested distance of the Oort cloud. While known KBOs are more likely to be
occulters, it is likely that  potential microlensing events, or combined microlens-occultation events, 
can be observed towards the Oort cloud.}
\label{fig:div}
\end{figure}

\section{Discussion}

We have explored the role of gravitational lensing 
when outer Solar system
objects, mainly members of the Kuiper belt and the Oort cloud, transit background
stars. The known population of KBOs, at distances
of order 40 AU with sizes of order few hundred kilometers and less, will 
always occult background stars.
Their occultation signature can be detected in monitoring campaigns involving
 few-meter class
telescopes with time sampling intervals of order few tenths of seconds (\cite{RoqMon00} 2000). While we have ignored here due to
our interest in large size objects, as discussed in \cite{RoqMon00} (2000), 
the detection rate of few kilometer size and below KBOs is partly enhanced by 
the diffraction effect that appears during the 
occultation. 

At distances much large than the currently known KBO population, the potential observability of a microlensing
event significantly increases. At distances of few 10,000 AU and more, corresponding to the Oort cloud,
objects with sizes of order hundreds of kilometers or more will gravitationally microlens background stars instead of simply occulting them. 
 For certain foreground object sizes, or mass, at favorable distances, 
one can potentially observe a combination of an occultation and a lensing event 
during the same transit. The detection of an onset of a lensing event on
an occultation light curve is interesting since the point at which the lensing
signature enters allows one to measure the ratio of the foreground source radius to its Einstein radius accurately
(\cite{Bro96} 1996). This additional information will aid in
constraining  physical parameters of the foreground population
beyond what is available solely from the occultation  or the
microlensing light curve. 

Note that the estimated KBO optical depth near 
the ecliptic is of order 10$^{-6}$  (\cite{RoqMon00} 2000); this is at 
the same level as the microlensing optical depth 
towards the galactic bulge. The Oort cloud optical depth, however, is
highly uncertain due to our limited knowledge on various properties of its 
population. The dynamical constraints, 
based on orbits of long-period comets, suggest a total
population of $\sim 10^{12}$ with a total 
mass of 38 $M_{\earth}$ (\cite{Wei96} 1996); 
these estimates are clearly uncertain for obvious reasons. While the total 
mass is
higher than that associated with KBOs, for the observable transit optical depth, what
is required is the distribution of source sizes. If sizes are all equal, 
then with a mass of order
$\sim$ 10$^{14}$ g, and radii of order few tenths of kilometers, Oort 
cloud will remain undetectable with observations that attempt to detect transists. 

The detectable transits, occultations and/or lensing, 
however, require the presence of objects
with masses of order $\sim$ 10$^{22}$ g or with radii of order few hundred 
kilometers.
Note that certain constraints on the KBO population limits the size 
distribution  of the outer  KBOs, at distances between 50 and 70 AU, to be 
below few hundred 
kilometers (\cite{Alletal01} 2001).  
Such surveys, however, are not sensitive to even massive objects with radii of thousands kilometers  at distances corresponding to Oort cloud 
suggesting that instead of direct detection techniques, such as through
imaging data, indirect techniques such as transit signatures will be needed to constrain its population.
Note that the monitoring of transits involving both KBOs and OCOs can be 
concurrently considered except that the
detection of events in monitoring data should involve the search for 
both a flux decrement as well as a possible increment due to a lensing
event. Note that any increment due to lensing can easily be
ascribed to massive bodies at 
large distances such that one breaks the usual degeneracy one encounters
in galactic microlensing studies involving the mass and distance of the object.
On the other hand, 
even if no lensing events are detected, any reliable upper limit on the
lensing optical depth towards the 
Oort cloud can be 
used to constrain the massive end of its population
and will aid in understanding the role 
OCOs play in the formation and evolution of the Solar system.

As we have discussed, 
the role of lensing on 
the occultation light curves of outer Solar system bodies is likely
to be only limited to distant Oort cloud members. The gravitational lensing 
effect, however, 
may already be important for objects in the inner Solar system.
For Pluto, at a distance of $\sim$ 39.5 AU and a mass of 0.002 $M_{\earth}$, 
the Einstein radius is of
order $\sim$ 15 km. This is small when compared to the Pluto radius of order 
1200 km and leads to the 
naive conclusion that any effects related to lensing by Pluto can be ignored 
when interpreting
its data. For precision calculations and parameter estimations, however, there
may be an additional consequence associated with lensing. 
While the magnification signature may not be dominant,
gravitational lensing also induce variations in astrometry, mainly a relative change in
the image position with respect to the unlensed position.
When interpreting light curves to derive atmospheric parameters, 
as was done in \cite{EllYou92} (1992),
it may  be necessary to account for the shift in image position due to 
lensing along with variations
arising from the atmospheric refraction effect.  If not accounted properly, 
one will wrongly conclude
the depth to which the light curve probes the Pluto's atmosphere with an 
error that is of the same order
 as the size of the Einstein radius. 

The best published data on an 
occultation by Pluto comes from
the 9th June 1998 event involving the background star P8 (\cite{Miletal93} 1993). 
At the inner most depths probed by refraction, the light curve associated with this event 
showed an anomalous gradient beyond a simple refractive atmosphere. This gradient has been modeled either
as extinction due to a haze layer or due to an abrupt thermal gradient. 
A preliminary calculation of the astrometric lensing correction to the light curve depth indicated that 
the abrupt change in the light curve was unlikely due to gravitational lensing modifications. 
Recently, it was reported that several new light curves related to an occultation by Pluto has now been
obtained. It'll be an interesting exercise to see if these data require an accounting of the
astrometric shift in the background source image during the inner depths of the occultation due to gravitational lensing.

To summarize, minor bodies of the outer Solar system will always occult background stellar
sources. There is still some 
possibility that a distance sample of objects, such as members of
the Oort cloud, will microlens background stars.
Another possibility is that there will be a combined signature 
of an occultation and a lensing event during the same transit.
These events, and occultation and lensing only events as well, 
can be extracted from continuous monitoring 
campaigns similar to those that
are currently pursued to detect galactic microlensing towards the bulge
and the Magellenic clouds.
The data sampling  intervals of a Solar system targeted campaign, however, should be at the order of few tenths of seconds and is within reach experimentally
in the near future.

\smallskip
{\it Acknowledgments:} This research was supported at Caltech by a senior research fellowship from the
Sherman Fairchild foundation and additional support from the Department of Energy.  The author thanks
Marc Kamionkowski for encouraging the author to work on topics beyond cosmology.

\end{document}